\documentclass[pre,aps,twocolumn,superscriptaddress,nofootinbib]{revtex4-2}
\pdfoutput=1
\usepackage{amsmath, amsthm, amssymb}
\usepackage{amsfonts}
\usepackage{graphicx}
\usepackage{dcolumn}
\usepackage{bm}
\usepackage{textcomp}
\usepackage[normalem]{ulem}

\usepackage[titletoc,title]{appendix}

\usepackage{epsfig}
\usepackage{subfigure}
\usepackage{url}
\usepackage{float}
\usepackage{cases}

\usepackage{colordvi}
\usepackage[usenames,dvipsnames]{xcolor}

\usepackage[colorlinks=true, urlcolor=blue, anchorcolor=blue, citecolor=blue,filecolor=blue,linkcolor=blue,menucolor=blue
]{hyperref}

\usepackage[titletoc,title]{appendix}

\begin{document}
\title{Speed fluctuations of a stochastic Huxley--Zel'dovich front}

\author{Evgeniy Khain}
\email{khain@oakland.edu}
\affiliation{Department of Physics, Oakland University, Rochester, MI 48309, USA}

\author{Baruch Meerson}
\email{meerson@mail.huji.ac.il}
\affiliation{Racah Institute of Physics, Hebrew University of
Jerusalem, Jerusalem 91904, Israel}

\author{Pavel V. Sasorov}
\email{pavel.sasorov@gmail.com}
\affiliation{ELI Beamlines Facility, Extreme Light Infrastructure ERIC, 252 41 Dolni Brezany,  Czech Republic}


\begin{abstract}

The empirical speed of travelling reaction--diffusion fronts fluctuates due to the intrinsic shot noise of the reactions and diffusion. Here we study the long-time front speed fluctuations of a stochastic Huxley--Zel'dovich front. It involves a population of particles $A$ which perform a fast continuous-time random walk on a one-dimensional lattice and undergo reversible  on-site reactions $2A \rightleftarrows 3A$.
This front describes an invasion of $A$-particles into an initially empty region of space which, in a deterministic description, is marginally stable but nonlinearly unstable with a zero instability threshold.
Typical fluctuations of this front can be described as front diffusion in a reference frame moving with the average front speed. According to the existing perturbation theory, the shot-noise-induced systematic shift of the average front speed, $\delta c$, and the front diffusion coefficient, $D_f$, are both expected to scale with $N$ as $1/N$, where $N \gg 1$ is the typical number of particles in the transition region. Furthermore, $D_f$ can be determined perturbatively in the small parameter $1/\sqrt{N}$. Our Monte Carlo simulations support these asymptotic results, but also reveal a long-lived anomalous behavior of the first few particles before they reach the expected diffusion asymptotic. We also study  large deviations of the empirical speed of the front at long times. These are dominated by optimal histories of the system in the form of a propagating front which travels with a speed different from the average speed, or even travel in the wrong direction.

\end{abstract}
\maketitle

\section{Introduction}
\label{intro}

Reaction-diffusion fronts are prone to intrinsic shot noise of the elemental processes of reactions and diffusion. One visible effect of the shot noise on macroscopic reaction-diffusion fronts is a small systematic shift of the average speed of the front compared with the prediction from deterministic theory. An important additional effect is fluctuations of the front speed around the mean. These effects attracted much interest across physics, chemistry, and biology, see Refs. \cite{vanSaarloos03,Panja} for reviews.

The small systematic shift of the average speed of a stationary macroscopic front strongly depends on whether the front propagates into a metastable or a linearly unstable region as predicted by deterministic theory of the fronts \cite{vanSaarloos03,Panja,Derrida1,MSK,Benguria}. The same is true for the typical -- that is small -- fluctuations of the front speed around the mean. These are commonly described in terms of a front diffusion in a reference frame moving with the average front speed, and this diffusion is characterized by a diffusion coefficient $D_f$ \cite{Panja}. For the fronts propagating into a metastable state $D_f$ scales with $N$ (where $N\gg 1$ is the typical number of particles in the transition region)  as $1/N$ \cite{Panja,MSK,KM}. The coefficient in this relation can be calculated by applying a regular perturbation expansion in the small parameter $1/\sqrt{N}\ll 1$  to the Langevin equation: a stochastic reaction-diffusion equation governing the front on macroscopic time and length scales \cite{MSK}.

For the fronts propagating into a linearly unstable state, the scaling behavior of the front diffusion coefficient $D_f$ with $N$ can be dramatically different. The \emph{pulled} fronts -- a subclass of fronts, propagating into a linearly unstable state -- are especially sensitive to the shot noise in their leading edge \cite{vanSaarloos03,Panja,Tsimring,Derrida1,Derrida06,Hallatschek}.  For such fronts (which, in the deterministic limit of $N\to \infty$, propagate with the linear spread velocity \cite{vanSaarloos03}),  $D_f$ scales with $N$ as $\ln^{-3} N$ and is therefore quite large \cite{Derrida06}. This remarkable amplification of the front  diffusion is caused by a few particles which outrun the front, multiply by branching, and ultimately dominate the fluctuations of the
front propagation speed \cite{Derrida06}.

The \emph{pushed} fronts are governed by the body of the front rather than by its leading edge, and their asymptotic speed in the deterministic limit exceeds the linear spread velocity  \cite{vanSaarloos03}.  As far as the effect of the shot noise on $D_f$ is concerned, the pushed fronts can be subdivided into two subgroups -- the strongly and the weakly pushed fronts -- depending on whether the integrals over space, which appear in the regular perturbation theory in $1/\sqrt{N}$ \cite{MSK}, converge or diverge, respectively \cite{Birzu2018,KMS}. It is known by now that, for strongly pushed fronts, and sufficiently far from
the transition point between the strongly and weakly pushed fronts, the simple leading-order scaling $D_f \sim 1/N$ continues to hold \cite{Birzu2018,KMS}.
These fronts, however, exhibit a striking anomaly: they behave diffusively only
when analyzed at extremely large time lags $\Delta t\gg N\gg 1$  \cite{KMS}.   This anomaly, which invalidates the Langevin equation at (macroscopic!) intermediate time lags $O(N)$,  also results from a few particles which outrun the front, multiply by branching, and cause large positive deviations of the front speed \cite{KMS}.

The  stochastic reaction-diffusion front which we consider in this work is special: it
propagates into a region which, in a deterministic description, is marginally stable. In a continuous deterministic theory, this front is governed by the Huxley-Zel'dovich (HZ) equation
\begin{equation}\label{Z}
    \partial_t u = u^2(1-u)+\partial_{x}^2 u\,,
\end{equation}
where the rate coefficients in the r.h.s. are all set to $1$ by a proper rescaling of the coordinate $x$, time $t$ and the density $u(x,t)$. Equation~(\ref{Z})  describes, in one spatial dimension, invasion of an initially empty region of space, $u=0$, by a population with density $u=1$. Here the empty state $u=0$ is linearly stable, but nonlinearly unstable -- with a zero instability threshold -- due to the $u^2$ term. This makes Eq.~(\ref{Z}) different from the better known Fisher-Kolmogorov-Petrovskii-Piscounov (FKPP) \cite{FKPP}
equation
\begin{equation}\label{Feq}
    \partial_t u = u(1-u)+\partial_{x}^2 u,
\end{equation}
where the empty state $u=0$ is linearly unstable, and which is a classical example of a pulled front \cite{vanSaarloos03}.

The deterministic equation (\ref{Z}) was introduced independently in two different areas of science. In theoretical biology, where it models the propagation of a new advantageous recessive gene in sexually reproducing populations, and the propagation of an electrical signal in a nervous system, it is called the Huxley equation \cite{Broadbridge2002,Broadbridge2004,Broadbridge2016}. In combustion theory, Eq.~(\ref{Z}) provides a simplified model of flame propagation due to heat diffusion and Arrhenius-like chemical reaction heat release \cite{Zel'dovichbook}, and it is known there under the name of Zel'dovich equation \cite{DMV,Gilding}.

When starting from an initial density $u(x,t=0)$ such that (i) $u(x\to -\infty,t=0) =1$ and (ii) $u(x\to +\infty,t=0) =0$, the solution of the deterministic HZ equation (\ref{Z}) converges at long times, $t\gg 1$, to a travelling front solution (TFS) $U(x-c t)$, where $c$ is the asymptotic front speed.  The function $U(\xi)$ obeys the ordinary differential equation (ODE)
\begin{equation}\label{MFeq}
    U^{\prime\prime}+c U^{\prime}+U^2-U^3=0
\end{equation}
and the boundary conditions $U(\xi \to -\infty)=1$ and $U(\xi \to \infty)=0$. Among the TFSs there is a special one, which is selected when the initial condition $u(x,t=0)$ falls off sufficiently fast at $x \to \infty$ (such as, for example, in the invasion of an empty region), see e.g. Ref. \cite{van Saarloos1}, Appendix C. This TFS, which is unique up to translations, falls off exponentially with $\xi$:
\begin{equation}\label{mffront}
U_0(\xi) =\frac{1}{1+\exp \left(\frac{\xi}{\sqrt{2}}\right)}\,,
\end{equation}
with $c=c_0=1/\sqrt{2}$. Since the uniform state $u=0$ is marginally stable, the HZ front is a pushed front.

Being interested in the shot-noise induced fluctuations of the front speed around its mean value, here we propose, in Sec. \ref{model}, a stochastic reaction-diffusion model -- the stochastic HZ model -- a coarse-grained deterministic  limit of which coincides with Eq.~(\ref{Z}). The model involves a single population of particles $A$ which perform continuous-time random walk on a one-dimensional lattice and undergo reversible on-site reactions, $2A \rightleftarrows 3A$. Assuming that the random walk is very fast, we can approximate it by a Brownian motion. As a result, typical macroscopic fluctuations in this system are described by a Langevin-type stochastic partial differential equation (PDE) for the coarse-grained particle density $u(x,t)$ \cite{MSK,KM}.  The ensuing \emph{stochastic} HZ front turns out to be strongly pushed, and we calculate, in Sec. \ref{Df}, the asymptotic front diffusion coefficient $D_f \sim 1/N$ by applying the linear perturbation theory in the small parameter $1/\sqrt{N} \ll 1$ \cite{MSK}. A systematic shift of the average speed of this front due to the fluctuations, $\delta c\equiv c-c_0$ is also expected to scale with $N$ as $\sim 1/N$.

In Sec. \ref{WKB} we study \emph{large deviations} of the empirical speed $c$ of a stationary front, as described by the probability distribution $\mathcal{P}(c,N,\Delta t)$, where both the number of particles in the transition region of the front $N$, and the observation time $\Delta t$, are assumed to be large.  To this end we employ the macroscopic fluctuation theory MFT, also called the optimal fluctuation method (OFM), the WKB (Wentzel-Kramers-Brillouin) approximation, the instanton theory, and the weak-noise theory. Combining exact, perturbative and numerical solutions of the MFT equations, we calculate the long-time large deviation function of the front speed, which has the following structure:
\begin{equation}\label{ldf}
- \ln \mathcal{P}(c,N,\Delta t) \simeq  N \nu \Delta t r(c),
\end{equation}
with a rate function $r(c)$.  (Here $\nu$ is a characteristic rate of the on-cite reactions which we will define shortly.) In particular, we observe that the rate function $r(c)$ exhibits qualitatively similar behaviors at $c<c_0$ and at $c>c_0$, as the leading-edge particles, outrunning the front, do not play a significant role here.

In Sec. \ref{MC} we compare the asymptotic formula for $D_f\sim 1/N$  and the scaling behavior $\delta c \sim 1/N$ as predicted in Sec. \ref{Df}, as well as the rate function $r(c)$ as predicted in Sec. \ref{WKB}, with the results of extensive Monte Carlo (MC) simulations of the microscopic stochastic HZ model. To this end we measure:
\begin{enumerate}

\item The empirical front speed,
\begin{equation}
\label{P130}
\bar{c}(t,\Delta t;N)=\frac{\left\langle X(t+\Delta t) -X(t)\right\rangle}{\Delta t}.
\end{equation}
\item The variance of the increment (or equivalently, the second-order structure function) of the front position $X(t)$,
\begin{eqnarray}\label{functionC}
\mathcal{V}(t, \Delta t;N) &=& \langle \left[X(t+\Delta t) -X(t)\right]^2 \rangle \nonumber \\
&-&\left\langle X(t+\Delta t)-X(t)\right\rangle^2.
\end{eqnarray}
\item The probability distribution function (PDF) of the increment $X(t+\Delta t) -X(t)$.
\end{enumerate}

At long times, $t\gg 1$, the empirical front speed $\bar{c}(t,\Delta t;N) \to c_*(N) $ becomes independent of $t$ and $\Delta t$; it depends only on $N$. The simulations show that $c_*(N)$ converges to the continuous deterministic speed $c_0 =1/\sqrt{2}$ of the HZ front at $N \to \infty$ according to the expected theoretical relation $\delta c \sim 1/N$, and we determine the coefficient in this relation, which turns out to be negative.

The variance $\mathcal{V}(t,\Delta t;N)\to V(\Delta t;N)$, as well as the PDF of the increment $X(t+\Delta t) -X(t)$, also become independent of $t$ at long times. Moreover, here $V(\Delta t;N)$ exhibits a linear dependence on $\Delta t$, which implies a standard diffusion behavior of the front. The simulations show that the ensuing diffusion coefficient
\begin{equation}\label{Dfdefinition}
D_f=\frac{\mathcal{V}(\Delta t;N)}{2\Delta t}
\end{equation}
is close to the theoretical prediction from Sec. \ref{Df}. We also observe that the convergence rate to this asymptotic value of $D_f$
becomes much longer, in terms of the lag time $\Delta t$, if we measure the front position $X$ by the position of one of a few leading particles, rather than by the position of a particle well inside the body of the front. For the leading particles the function $\mathcal{V}(\Delta t)$ is strongly sub-linear at intermediate values of $\Delta t$, implying an anti-persistent, sub-diffusive behavior at these time scales.

Our results are briefly summarized and discussed in Sec. \ref{discussion}.

\section{Stochastic Huxley-Zel'dovich model}
\label{model}

Our stochastic HZ model is defined on a one-dimensional lattice with the lattice spacing $h$. The particles  perform continuous-time symmetric random walk between neighboring sites with (per particle) hopping rate $D_0$,  and also undergo the on-site reaction $2A\rightarrow 3A$ with rate constant $\beta$ and the inverse reaction $3A\rightarrow 2A$ with rate constant $\sigma$. In the absence of random walk,  the on-site master equation for the probability $P_n(t)$ of observing $n$ particles on a given site reads
\begin{equation}\label{onsitemaster}
\!\dot{P}_n(t)\!=\!\lambda_{n-1} P_{n-1}(t) +\mu_{n+1}  P_{n+1}(t)-(\lambda_n+\mu_n)  P_n(t),
\end{equation}
with combinatorial rates $\lambda_n=\beta n(n-1)/2$  and $\mu_n=\sigma n(n-1)(n-2)/6$.
In the presence of very large hopping rate $D_0$ -- see Eq.~(\ref{largeD}) below for the required strong inequality -- the random walk rapidly establishes a Poisson distribution
of the number of particles on each site. As a result, the (much smaller) on-site particle birth and death rates are well approximated by their mean-field values
$\lambda(n) = \beta n^2/2$ and $\mu(n) = \sigma n^3/6$, respectively. (Note that $n$ here is a macroscopic density which does not have to be integer.) The nontrivial deterministic fixed point (the carrying capacity) of a constant-density population is therefore at  $n= K \equiv 3\beta/\sigma$.
Introducing the rescaled population size $u=n/K$ and the effective rate constant $\nu \equiv 3 \beta^2/(2\sigma)=\beta K/2$, we can rewrite the birth and death rates as
\begin{equation}\label{birthdeathratesq}
\lambda(u) = \nu u^2  \quad \text{and}\quad \mu(u) = \nu u^3 \,.
\end{equation}
We define the diffusion coefficient  in a standard way: $D=D_0h^2/2$. Our crucial assumption in this work is that the characteristic diffusion length $\ell_D=(D/\nu)^{1/2}$  is much larger than the lattice spacing $h$. In terms of the process rates, this strong inequality reads
\begin{equation}\label{largeD}
\frac{D_0}{\beta K} \gg 1.
\end{equation}
In this limit the symmetric random walk can be approximated by a Brownian motion with the diffusion constant $D$.
We also demand that the typical number of particles over the characteristic diffusion length $\ell_D$ be very large:  $N=K\ell_D/h \gg 1$.  Using this definition of $N$, we can rewrite the strong inequality (\ref{largeD}) as $N \gg K$. This condition and the condition $N\gg 1$ can be written as a single condition,
\begin{equation}\label{NandK}
N \gg \text{max} \left(K,1 \right)\,.
\end{equation}
Once this condition is met, we arrive at a continuous-in-space  macroscopic description of the system in terms of the coarse-grained particle density $u(x,t)$ \cite{macroscopic}. Taking the deterministic limit, one arrives at
the HZ equation~(\ref{Z}).

We also note here that, irrespectively of the assumption (\ref{NandK}), the rescaled stochastic HZ model on the lattice is completely determined by the two dimensionless parameters $N$ and $K$.

\section{Typical fluctuations of the front speed}
\label{Df}

By analogy with fronts, propagating into a metastable state \cite{MSK,KM}, we will assume (and then verify this assumption in MC simulations)
that  typical macroscopic fluctuations \cite{macroscopic} of the HZ front can be described by a Langevin equation -- a stochastic PDE, which deterministic part coincides with the HZ equation (\ref{Z}), and which has two  effective noise terms corresponding to the reactions and diffusion, respectively.  In the rescaled variables $\nu t \to t$ and $x/\ell_D \to x$ this equation has the form  \cite{MSK}
\begin{eqnarray}\label{P040}
\partial_t u&=&f(u)+\partial_x^2 u \nonumber \\
&+&\frac{1}{\sqrt{N}}\left[\sqrt{g(u)}\eta(x,t)+\partial_x\left(\sqrt{2u}\chi(x,t)\right)\right]\, ,
\end{eqnarray}
where
\begin{equation}\label{P030}
f(u)=u^2-u^3\quad \text{and} \quad g(u)=u^2+u^3\,,
\end{equation}
$\eta(x,t)$ and $\chi(x,t)$ are independent centered Gaussian noises, delta-correlated both in $x$ and
in $t$. The nonconservative noise term proportional to $\eta(x,t)$) in the r.h.s. of Eq.~(\ref{P040}) describes, on the macroscopic scale, the noise of the reactions $2A\rightleftarrows 3A$. The conservative noise term proportional to $\chi(x,t)$ comes from the noise of random walk.
Importantly, the dimensionless parameter $K$ does not appear in the Langevin equation (\ref{P040}).

The small parameter $1/\sqrt{N}\ll 1$ in the r.h.s. of Eq.~(\ref{P040}) calls for a perturbation theory around the deterministic travelling front
solution (\ref{mffront}). For the fronts propagating into a metastable state, such a perturbation  theory was developed in Ref. \cite{MSK}, although similar approaches had been used earlier
for phenomenological stochastic travelling front models \cite{Mikhailov,Pasquale,Rocco}. More recently, this perturbation theory was extended
to the strongly pushed fronts propagating into an unstable state \cite{KMS,Birzu2018}.

The stochastic HZ front is exactly at the boundary between the fronts propagating into a metastable state and the strongly pushed fronts propagating into an unstable state. This ``very strongly" pushed front can therefore be described by the standard perturbation theory \cite{MSK}.
In the first order in $1/\sqrt{N}$ one can calculate the diffusion coefficient of the front $D_f$, which is predicted to scale with $N$ as $1/N$.   The systematic shift of the front speed $\delta c$, which also scales with $N$ as $1/N$ appears only in the second order of this perturbation theory. As of present, such explicit second-order calculations -- which should give the numerical coefficient in this scaling relation -- have not been performed yet. Therefore, here we present a complete leading-order analysis for $D_f$, but limit ourselves to a numerical verification of the  $1/N$ dependence of $\delta c$ \cite{discreteness}.

The leading-order perturbative calculation of $D_f$ reduces to evaluation of integrals over $\xi$ of the unperturbed deterministic front solution $U_0(\xi)$, as described by Eq.~(\ref{mffront}), its $\xi$-derivative $U_0'(\xi)$, and the function $g[U_0(\xi)]$, where $g(u)$ is defined in Eq.~(\ref{P030}).  According to Eq.~(17) from Ref.~\cite{MSK}, and restoring the dimensional quantities,
\begin{equation}\label{P060}
D_f\simeq \frac{J_1+J_2}{J_3^2} \frac{D}{N}\,,
\end{equation}
where
\begin{eqnarray}
\label{J123}
  J_1 &=& \frac{1}{2}\int\limits_{-\infty}^\infty \left[U_0^\prime(\xi)\right]^2 e^{2c_0\xi} g\left[U_0(\xi)\right]\, d\xi\,,\nonumber\\
  J_2 &=& \int\limits_{-\infty}^\infty U_0(\xi)  \left[\left(U_0^\prime(\xi) e^{c_0\xi}\right)^\prime\right]^2\, d\xi\,,\nonumber\\
  J_3 &=& \int\limits_{-\infty}^\infty \left[U_0^\prime(\xi)\right]^2 e^{c_0\xi}\, d\xi\,,
\end{eqnarray}
and $U_0(\xi)$ is given by Eq.~(\ref{mffront}). To remind the reader, $c_0=1/\sqrt{2}$, while the diffusion coefficient $D=D_0 h^2/2$, where $h$ is the lattice spacing.   All the integrals in Eq.~(\ref{J123}) can be evaluated analytically, and we obtain
\begin{equation}\label{J123results}
J_1 = J_2 = \frac{1}{30\sqrt{2}}\,,\quad J_3 =\frac{1}{3\sqrt{2}}\,.
\end{equation}
Using Eqs. (\ref{P060}) and (\ref{J123results}), we arrive at the following prediction for the diffusion coefficient of the HZ front:
\begin{equation}\label{Dfpredicted}
D_f \simeq \frac{3\sqrt{2}}{5}\, \frac{D}{N} = 0.8485\dots \frac{D}{N}\,.
\end{equation}
As we discussed above, the systematic shift of the average speed of the HZ front should also behave, in the leading order, as $1/N$:
\begin{equation}\label{deltacprediction}
\delta c \simeq \frac{\text{const}}{N}\,,
\end{equation}
where the constant factor (and even its sign) are presently unknown. In Sec. \ref{MC} we will test the predictions (\ref{Dfpredicted}) and (\ref{deltacprediction}) in MC simulations.
We now turn to a theoretical description of large deviations of the empirical front speed.

\section{Large deviations of the front speed}
\label{WKB}

\subsection{MFT equations}
\label{MFTeq}

For a pushed front to move with an average speed different from $c_0=1/\sqrt{2}$ during the time $\Delta t\gtrsim 1$, a significant and long-lived fluctuation-induced modification of the front density profile is needed. Such a modification involves many particles. As a result, the probability density of the front speed is exponentially small in the characteristic number of particle in the transition region, $N\gg 1$. Using the large parameter $N\gg 1$, we
can apply the macroscopic fluctuation theory  (MFT)
\cite{MFT}, extended to account for on-site reactions among particles, see Refs. \cite{Jona-Lasinio1,Jona-Lasinio2,Bodineau,EK,MS2011}.
The MFT demands the same strong inequality (\ref{NandK}) as the Langevin description. An additional assumption of the MFT
is that the probability distribution of the process of interest, conditional on a large deviation, is dominated by the optimal -- that is most likely -- history of this process.
The optimal history is described by Hamilton's equations for the optimal density
$u(x,t)$ and the ``conjugate momentum" density $p(x,t)$.  For our reactions $2A \rightleftarrows 3A$ the MFT equations,
in the rescaled units of Eq.~(\ref{P040}),  read \cite{Jona-Lasinio1,Jona-Lasinio2,Bodineau,EK,MS2011}:
\begin{eqnarray}
\!\!\!\partial_t u&=&u^2 e^p - u^3 e^{-p}+ \partial_x^2u -2\partial_x\left(u\partial_x p\right),
\label{p100}\\
\!\!\!\partial_t p&=& -2u(e^p-1) -3u^2 (e^{-p}-1)-\partial_x^2 p -\left(\partial_x p\right)^2.
\label{p110}
\end{eqnarray}
The corresponding Hamiltonian is $\int dx\, w$, with the Hamiltonian density
\begin{equation}\label{hamdensity}
w=u^2 (e^{p}-1) +u^3 (e^{-p}-1) -\partial_x u\, \partial_x p+ u (\partial_x p)^2\,.
\end{equation}
The first two terms of Eq.~(\ref{hamdensity}) describe the fluctuations coming from the on-site reactions $2A \rightleftarrows 3A$, whereas the last two terms describe fluctuations of the Brownian motion. A straightforward way to derive the reaction terms employs an on-site WKB approximation, which is based on the assumption that the optimal number of particles on each relevant site is much larger than unity, hence the demand $K\gg 1$ \cite{EK,MS2011}. Under condition (\ref{NandK}), however, the strong inequality $K\gg 1$ becomes unnecessary. The results of our MC simulations, presented in Fig. \ref{figPDFsim} below, support this observation.

The motion of the system (\ref{p100}) and (\ref{p110}) on the zero-energy invariant manifold $p=0$ is described by the deterministic HZ equation~(\ref{Z}).

The boundary conditions at minus infinity, $u(x=-\infty,t)=1$ and
$p(x=-\infty,t)=0$, correspond to the fixed point $(u=1,p=0)$ of the \emph{on-site} Hamiltonian,
\begin{equation}\label{Honsite}
H_0(u,p)= u^2 (e^p-1) +u^3 (e^{-p}-1)\,.
\end{equation}
Front propagation into an empty region corresponds to the boundary conditions $u(x=-\infty,t)=1$ and $u(x=+\infty,t)=0$. The momentum density $p(x,t)$ must be bounded at finite $x$, but it can be unbounded, $p(x\to +\infty,t) = -\infty$, as in the case of the fluctuating FKPP front \cite{MSfisher}.

Having found the optimal history for the transition between an initial condition $u_1(x,t=t_1)$ and a final condition $u_2(x,t=t_2)$, where $t_2=t_1+\Delta t$, we can evaluate, up to a pre-exponential factor,  the probability density of this transition:
\begin{eqnarray}
   &-&\ln \mathcal{P}[u(x,t_1)\to u(x,t_2);N] \nonumber \\
   &\simeq&S[u(x,t_1)\to u(x,t_2);N] \nonumber \\
   &=&\nu N \int_{-\infty}^{\infty} dx \int_{t_1}^{t_2} dt \left[p(x,t) \partial_t u-w\right].
  \label{action0}
\end{eqnarray}

The MFT equations become somewhat simpler if we make the canonical Hopf-Cole transformation from the variables $u$ and $p$ to the new variables
$\mathcal{Q}=u e^{-p}$
and $\mathcal{P}=e^p-1$ (the shift by $1$ in $\mathcal{P}$ preserves the mean-field hyperplane
at $\mathcal{P}$=0). Obviously, $\mathcal{Q}\geq 0$ and  $\mathcal{P}\geq -1$. The
generating function of this canonical transformation can be chosen to be $\int dx \, F(u,\mathcal{Q})$, where
\begin{equation}\label{genf}
    F(u,\mathcal{Q})=u\,\ln \frac{u}{\mathcal{Q}}-u+\mathcal{Q}\,.
\end{equation}
The new Hamiltonian density becomes
\begin{equation}\label{density}
\mathcal{W}=\mathcal{Q}^2 (1-Q)\mathcal{P}(1+\mathcal{P})^2 - \partial_x \mathcal{Q}\, \partial_x \mathcal{P}\,.
\end{equation}
The zero-energy lines of the \emph{on-site}
Hamiltonian $\mathcal{H}_0(\mathcal{Q},\mathcal{P})=
\mathcal{Q}^2 (1-Q)\mathcal{P}(1+\mathcal{P})^2$ in the phase plane $(\mathcal{Q},\mathcal{P})$ form a rectangle. In the new variables, the Hamilton's equations are
\begin{eqnarray}
  \partial_t \mathcal{Q}  &=& \mathcal{Q}^2(1-\mathcal{Q})(1+\mathcal{P})(1+3\mathcal{P})+\partial_x^2 \mathcal{Q}\,, \label{1B}\\
  \partial_t \mathcal{P} &=& \mathcal{Q}(3\mathcal{Q}-2)\mathcal{P}(1+\mathcal{P})^2 -\partial_x^2 \mathcal{P}\,. \label{2B}
\end{eqnarray}

The zero-energy  invariant manifold $\mathcal{P}(x,t)=0$ describes the deterministic, or mean-field, dynamics, that is the formation and propagation of the asymptotic deterministic TFs (\ref{mffront}). There is also the \emph{equilibrium} manifold $\mathcal{Q}=1$ which arises in this problem because of the reversibility of the reactions $2A \rightleftarrows 3A$, cf. Ref. \cite{MSfisher}. As we will see shortly, however, this manifold is inaccessible.

In the new variables, Eq.~(\ref{action0}) becomes
\begin{eqnarray}
   -\ln \mathcal{P}(c,\Delta t,N)&\simeq& S = N \nu \int_{-\infty}^{\infty} dx \,\Big\{ \Delta F(u,\mathcal{Q}) \nonumber \\
   &+&\int_0^{\Delta t} dt
  \left[\mathcal{P}(x,t) \partial_t \mathcal{Q}-\mathcal{W}\right]\Big\}\,,
  \label{action}
\end{eqnarray}
where
\begin{equation}\label{increment}
\Delta F \equiv F(u,Q)|_{t=\Delta t}-F(u,Q)|_{t=0}
\end{equation}
is the increment of the generating function density~(\ref{genf}) of the Hopf-Cole transformation between the initial time $t=0$ and the final time $t=\Delta t$, and $\mathcal{W}$ was defined in Eq.~(\ref{density}).

Now suppose that $\Delta t =t_2 -t_1 \gg 1$.    Similarly to other problems of  large-deviation statistics of front speed \cite{MSK,MSfisher,MVS,MS2024}, a natural simplifying assumption in this case is that the optimal history, conditional on $c \neq c_0$, is described, in the leading order in $\Delta t\gg 1$, by a continuous family of \emph{fluctuating TFSs} $\mathcal{Q}=\mathcal{Q}(x-ct)$ and $\mathcal{P}=\mathcal{P}(x-ct)$ of the MFT equations~(\ref{1B}) and (\ref{2B}). (For negative large deviations of the speed of \emph{pulled} fronts this assumption
was verified via numerical solutions of the complete time-dependent MFT problem  \cite{MVS}.)

The functions $\mathcal{Q}(\xi)$ and $\mathcal{P}(\xi)$ satisfy two coupled nonlinear ODEs:
\begin{eqnarray}
 \mathcal{Q}^{\prime\prime}+c \mathcal{Q}^{\prime}
 + \mathcal{Q}^2 (1-\mathcal{Q})(1+\mathcal{P})(1+3\mathcal{P}) &=&0\,, \label{ODEQ}\\
 \mathcal{P}^{\prime\prime}-c \mathcal{P}^{\prime}
 -\mathcal{Q} (3\mathcal{Q}-2) \mathcal{P} (1+\mathcal{P})^2 &=&0\,, \label{ODEP}
\end{eqnarray}
where the primes denote the $\xi$-derivatives. By virtue of the boundary conditions for $u$ and $p$ at $x=-\infty$, we demand that $\mathcal{Q}(\xi=-\infty)=1$ and $\mathcal{P}(\xi=-\infty)=0$. The values of $\mathcal{P}(\xi=\infty)$ and $\mathcal{Q}(\xi=\infty)$  depend on $c$.

Importantly,  Eqs.~(\ref{ODEQ}) and (\ref{ODEP}) possess a conservation law:
\begin{equation}\label{conservation}
\mathcal{H}_0[\mathcal{Q}(\xi),\mathcal{P}(\xi)]+\mathcal{Q}^{\prime} \mathcal{P}^{\prime}=\text{const}\,,
\end{equation}
where the constant in the r.h.s. vanishes because of the boundary conditions.

For a TFS with a specified speed $c$, Eq.~(\ref{action}) takes the form $ -\ln \mathcal{P}(c,\Delta t,N)\simeq N \Delta t \,\nu r(c)$, with the rate function
\begin{equation}\label{ratefunction}
r(c) = \dot{s}_1+\dot{s}_2\,.
\end{equation}
The first term $\dot{s}_1=\Delta F/\Delta t$ of the rate function can be viewed as the rate of the increment of the generating function $F$ between times $t=t_1$ and $t=t_2$, see Eqs.~(\ref{action}) and (\ref{increment}). The second term $\dot{s}_2$
comes from the second term in Eq.~(\ref{action}), and it can be written as
\begin{eqnarray}\label{accumrate1}
   \dot{s}_2&=&-\int_{-\infty}^{\infty} d\xi (c \mathcal{P} \mathcal{Q}^{\prime}+\mathcal{W})  \nonumber\\
   &=&\int_{-\infty}^{\infty} d\xi
    \left(\mathcal{P} \frac{\partial \mathcal{H}_0}{\partial \mathcal{P}}-\mathcal{H}_0\right) \nonumber\\
    &=& 2 \int_{-\infty}^{\infty} d\xi \, (\mathcal{Q}^2-\mathcal{Q}^3)(\mathcal{P}^2+\mathcal{P}^3)\,,
\end{eqnarray}
where we have used Eq.~(\ref{ODEQ}).

\subsection{MFT: Analytical results}

Although the MFT problem,  described by the two coupled nonlinear ODEs (\ref{ODEQ}) and (\ref{ODEP}), is simpler than the original MFT  problem, described by the PDEs (\ref{1B}) and (\ref{2B}), it still cannot be solved exactly for all $c\neq c_0$.  In this subsection we present some exact and perturbative analytical results that we obtained. Later on we will supplement the analytical solution by numerical ones.

\subsubsection{Fluctuation theorem}

One important special property of the reversible system of reactions and diffusion in the stochastic HZ model is a remarkable symmetry between $Q$ and $P$ for counter-propagating fronts with the same
absolute value of the speed $|c|$. Indeed, let us suppose that $\mathcal{Q}_1(x,t)$
and $\mathcal{P}_{1}(x,t)$ satisfy Eqs.~(\ref{1B})
and (\ref{2B}) and the boundary conditions  $\mathcal{Q}(x=-\infty,t)=0$ and
$\mathcal{P}(x=-\infty,t)=-1$. Then, as one can check, the functions $\mathcal{Q}_2=\mathcal{P}_1(x,-t)+1$
and $\mathcal{P}_2=\mathcal{Q}_1(x,-t)-1$ also satisfy Eqs.~(\ref{1B})
and (\ref{2B}) and the boundary conditions.
An important consequence of this symmetry is the following relation between the fluctuating TFSs
with speeds $c$ and $-c$:
\begin{equation}\label{symm}
    \mathcal{Q}_{(-c)}(\xi)=\mathcal{P}_{(c)}(\xi+C)+1\,,\;\mathcal{P}_{(-c)}(\xi)=
    \mathcal{Q}_{(c)}(\xi+C)-1\,,
\end{equation}
where $C$ is an aribitrary constant. In the variables $u$ and $p$ we obtain $u_{-c}(\xi) = u_c (\xi+ C)$. This
reversibility property -- which is a direct consequence of the detailed balance property of the microscopic model --
leads to a remarkable relation which has the form of a fluctuation theorem:
\begin{equation}\label{accumrateminusfinal}
r(-c)=r(c)+c\,.
\end{equation}
Not surprisingly, the same fluctuation theorem holds for the stochastic FKPP front in the special case when the linear and nonlinear reaction terms come from the reversible reactions $A\rightleftarrows 2A$ \cite{MSfisher}.

A simple but nontrivial consequence from Eq.~(\ref{accumrateminusfinal}) is the following exact result: $r(-c_0)=c_0$.

\subsubsection{Standing front: $c=0$}

\begin{figure}[ht]
\includegraphics[width=3.0 in,clip=]{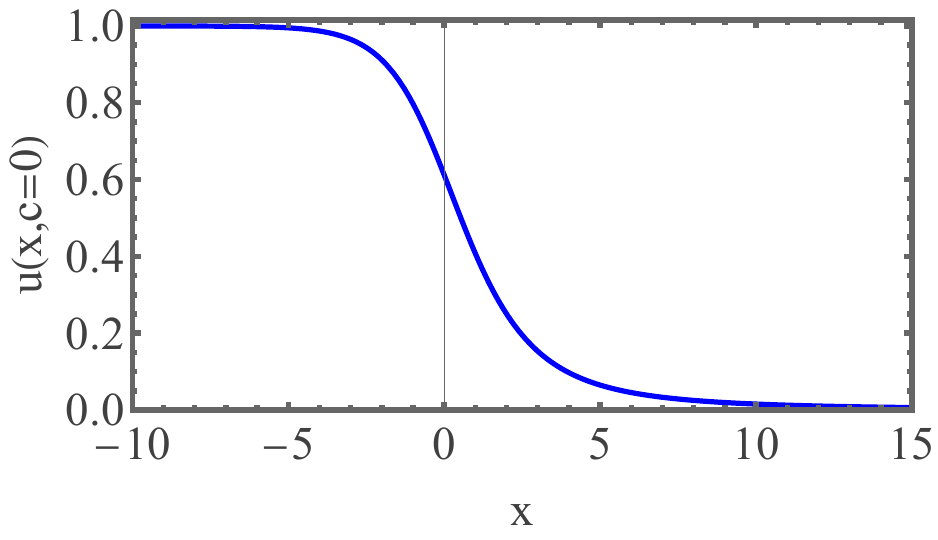}
\caption{The optimal density $u(x)$  for the standing noisy front, $c=0$, as described by the exact solution~(\ref{qzeroc}).}
\label{0}
\end{figure}

For $c=0$ the front motion is completely arrested, by a very strong and persistent fluctuation, during the whole time
$\Delta t$. In this case Eqs.~(\ref{ODEQ}) and (\ref{ODEP}), with the boundary conditions $\mathcal{Q}(x=-\infty)=1$, $ \mathcal{P}(x=-\infty)=0$, $\mathcal{Q}(x=\infty)=0$ and $\mathcal{P}(x=\infty)=-1$, can be solved exactly.
Here the symmetry relations (\ref{symm}) reduce to a single relation
to $\mathcal{P}=\mathcal{Q}-1$. Further, as one can check, for $c=0$ \emph{each} of the two equations (\ref{ODEQ}) and (\ref{ODEP}) takes the form
\begin{equation}\label{c0Q}
\mathcal{Q}^{\prime\prime}  + (\mathcal{Q}^2-\mathcal{Q}^3)(3\mathcal{Q}^2-2\mathcal{Q})=0\,.
\end{equation}
Integrating this equation once, we obtain
\begin{equation}\label{firstorder}
\mathcal{Q}^{\prime}+\mathcal{Q}^2-\mathcal{Q}^3=0,
\end{equation}
where the arbitrary constant vanishes because of the boundary conditions.
The solution of Eq.~(\ref{firstorder}) is elementary in terms of $x=x(\mathcal{Q})$:
\begin{equation}\label{c=0}
x=\frac{1}{\mathcal{Q}}+\ln \left(\frac{1}{\mathcal{Q}}-1\right),
\end{equation}
where we have arbitrarily fixed the front position.  In the original variables the optimal particle density $u(x)$ is described by
the relation
\begin{equation}\label{qzeroc}
x=\frac{1}{\sqrt{u}}+\ln \left(\frac{1}{\sqrt{u}}-1\right)\,.
\end{equation}
This density profile is shown in Fig. \ref{0}. At $x\gg 1$ $u(x)$ falls off as $u(x)\simeq x^{-2}$.

Now we can calculate the value of the rate function $r(c)$ at $c=0$ using the relation $\mathcal{P}(x)=\mathcal{Q}(x)-1$, Eq.~(\ref{accumrate1}),  and the fact that $\dot{s}_1=0$.  We obtain
\begin{equation}\label{sdotc0}
r(c=0)=\dot{s}_2(c=0)=2 \int_{-\infty}^{\infty} dx\,\mathcal{Q}^3 (1-\mathcal{Q})^3\,,
\end{equation}
where $\mathcal{Q}(x)$ is given implicitly by Eq.~(\ref{c=0}). Going over from the integration over $x$ to integration over  $\mathcal{Q}$, and using Eq.~(\ref{c=0}), we obtain after a simple algebra
\begin{equation}\label{dsdotc0result}
r(c=0)=2 \int_1^0 dQ \,\frac{dx}{dQ} \,\mathcal{Q}^3 (1-\mathcal{Q})^3=\frac{1}{6}\,,
\end{equation}
therefore $\mathcal{P}(c=0)\sim \exp(-N\nu \Delta t/6)$.

\subsubsection{$|c-c_0|\ll 1$: linear theory}
\label{pert}

The region of typical, Gaussian fluctuations of the front speed, $|c-c_0|\ll 1$, determines the diffusion coefficient of the front which we considered in Sec.  \ref{Df}. There is a major difference in this respect between the HZ front and the pulled fronts such as the stochastic FKPP front. For the pulled fronts, the front diffusion coefficient scales with $N$ as $1/\ln^3 N$ \cite{Derrida06}. Since it is contributed by a few particles, outrunning the front \cite{Derrida06}, its calculation is beyond the reach of a macroscopic theory such as the Langevin equation or the MFT \cite{MSfisher}.  For the (very strongly pushed) HZ front, one can probe the diffusion behavior of the front within the MFT. Furthermore, within the MFT framework, it can be found perturbatively in the small parameter $|c-c_0|$  \cite{MSK}. In the leading, linear order this calculation gives  a quadratic asymptotic  of the MFT rate function $r(c)=\dot{s}_2$ near $c=c_0$, which captures the Gaussian fluctuations.

The specific set of the on-site reactions, considered in Ref. \cite{MSK} as an example of the linear theory, included, in addition to the reactions $2A \rightleftarrows 3A$, also the linear decay $A\to 0$ which renders linear stability to the empty state. Adapting the results of Ref. \cite{MSK} to the fluctuating HZ equation, one can simply send the decay rate to zero. In the notation of Ref. \cite{MSK},  this corresponds to taking the limit $\delta\to 1$, leading to
\begin{equation}\label{reducedaction}
  r(c)\simeq f_G(c)\equiv \frac{5\sqrt{2}}{24} (c-c_0)^2\,,\quad |c-c_0|\ll c_0\,.
\end{equation}
The quadratic asymptotic~(\ref{reducedaction}) perfectly agrees with Eq.~(\ref{Dfpredicted}). Indeed,  Eq.~(\ref{Dfpredicted}) implies a Gaussian distribution of the fluctuations of the displacement of front $X$ relative to the mean:
\begin{equation}\label{gaussianX}
\mathcal{P}(\Delta X,t) \sim \exp\left(-\frac{\Delta X^2}{4 D_f \Delta t}\right),
\end{equation}
Plugging here $D_f$ from  Eq.~(\ref{Dfpredicted}), replacing $\Delta X$ by $(c_0-c)\Delta t$ and setting $D=1$, we arrive at
\begin{equation}\label{linMFT}
-\ln \mathcal{P} \simeq N \nu \Delta t f_G(c),
\end{equation}
as to be expected.

\subsubsection{The equilibrium manifold $\mathcal{Q}=1$ is inaccessible}

As in the case of the FKPP front with the reversible reactions $A\rightleftarrows 2A$ \cite{MSfisher}, there is an additional invariant manifold, $\mathcal{Q}(x,t) =1$, which solves the time-dependent Hamilton's equation  (\ref{1B}). In the original variables $u$ and $p$ this solution,  $p=\ln u$, describes an invariant
equilibrium manifold of Eqs.~(\ref{p100}) and (\ref{p110}). Eliminating $p$ in favor of $u$, one can see that each of the time-dependent equations (\ref{p100}) and (\ref{p110}) in this case becomes
\begin{equation}\label{timereversal}
\partial_t u =-\left(u^2-u^3+\partial_{x}^2 u\right)\,,
\end{equation}
which is the \emph{time-reversed} deterministic HZ equation  (\ref{Z}). Equation (\ref{timereversal}), subject to the boundary conditions $u(-\infty)=1$ and $u(\infty)=0$, describes an extremely large deviation when the front rapidly
propagates, with  $c\leq -c_0$, ``in the wrong direction". Further, there is a simple relation, $u(x-ct)=U_0(x+ct)$, between the family of these fronts and the family of \emph{deterministic} fronts $U(\xi)$ with $c\geq c_0=1/\sqrt{2}$. It is crucial, however, that for $c<-c_0$ the leading edge of these fluctuating fronts decays too slowly to be normalizable: $u(\xi \to -\infty)\sim - c/\xi$. As a result, these TFSs cannot emerge in the full time-dependent MFT problem, once we start from an initial condition such that $u(x>0,t=0)=0$. This stays in contrast with the case of the FKPP front with the reversible reactions  $A\rightleftarrows 2A$, where the TFSs with $c<-c_0$ have \emph{exponentially} decaying leading edges and therefore provide legitimate long-time optimal solutions in this regime \cite{MSfisher}.

\subsection{MFT: numerical results}
\label{nonlinearWKB}

For $c\neq c_0$ the solutions can be found numerically by a shooting method  (see e.g., Ref. \cite{shooting}), which effectively replaces the boundary value problem for Eqs.~(\ref{ODEQ}) and (\ref{ODEP}) on a sufficiently large but finite interval $|\xi|<L$ by
an initial value problem with an a priori unknown shooting parameter specified at either $x = -L$ \cite{MSK,MSfisher}, or $x=L$ (where, upon linearizing Eqs.~(\ref{ODEQ}) and (\ref{ODEP}) around the corresponding fixed points, one is left with a single shooting parameter). This parameter is iterated so as to avoid unphysical divergences of solutions to $\pm \infty$ within the numerical interval $|\xi|<L$. The accuracy of the method was verified, in particular, by comparing the numerical solution with the exact solution (\ref{c=0}) for $c=0$.

\begin{figure}[ht]
\includegraphics[width=2.0 in,clip=]{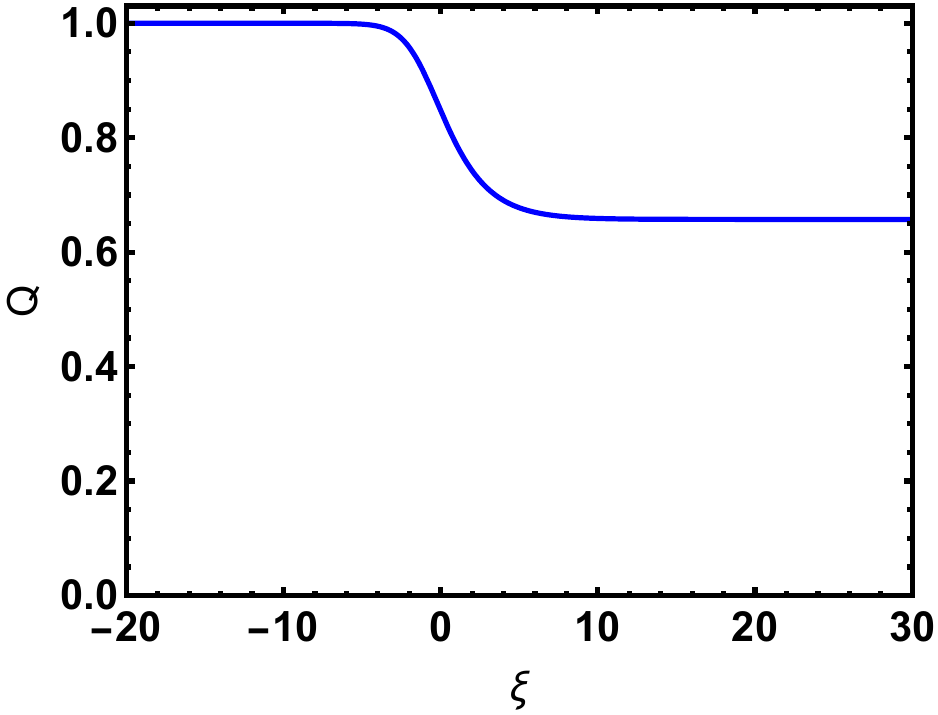}
\includegraphics[width=2.0 in,clip=]{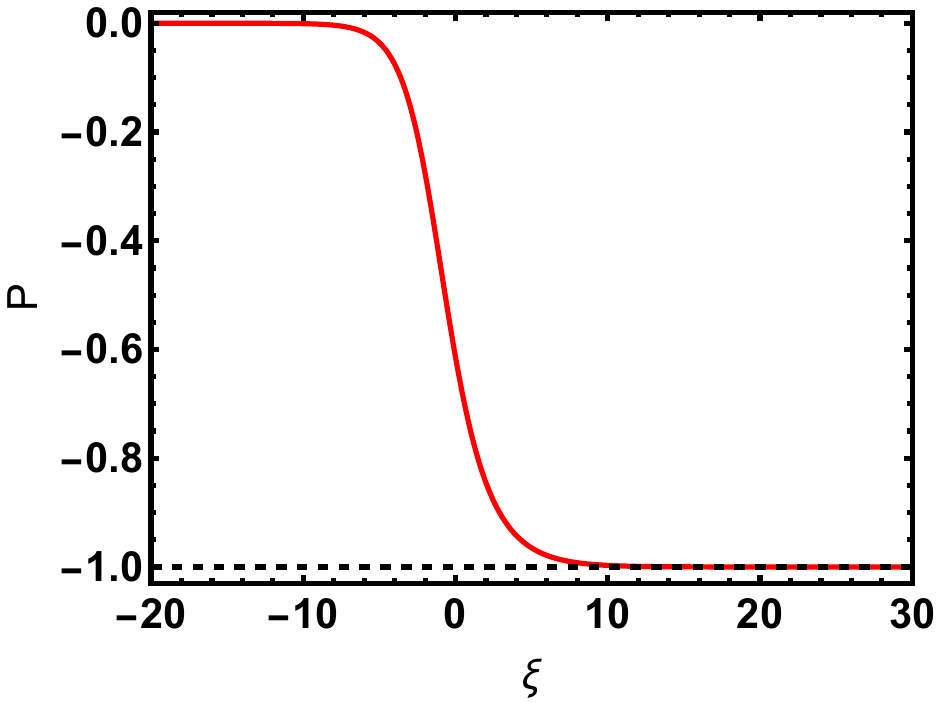}
\includegraphics[width=2.0 in,clip=]{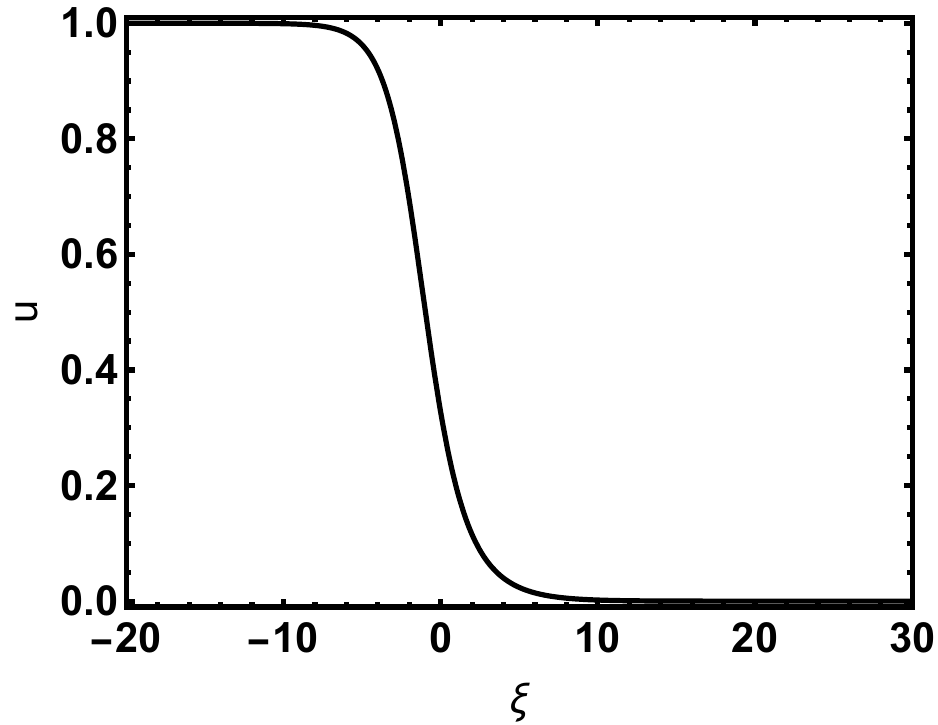}
\caption{The TFS found numerically for $c=-0.5$: $Q(\xi)$ (top left), $P(\xi)$ (top right), and $u(\xi) = Q(\xi) [1+P(\xi)]$ (bottom).}
\label{MFTsol}
\end{figure}

The resulting numerical solutions show that $Q(\xi = \infty) =0$, while $P(\xi = \infty)$ is negative and varies between $-1$ and $0$ for $0<c<c_0$, and it is positive for $c>c_0$. For $-c_0<c<0$,  $Q(\xi = \infty)$ varies between $0$ and $1$, while $P(\xi = \infty)=-1$. The whole regime $c<-c_0$ corresponds to the equilibrium manifold $Q=1$; here $P(\xi =\infty) = -1$.  Figure \ref{MFTsol} shows an example of the numerically found TFS, where $c=-0.5$.

\begin{figure}[ht]
\includegraphics[width=2.0 in,clip=]{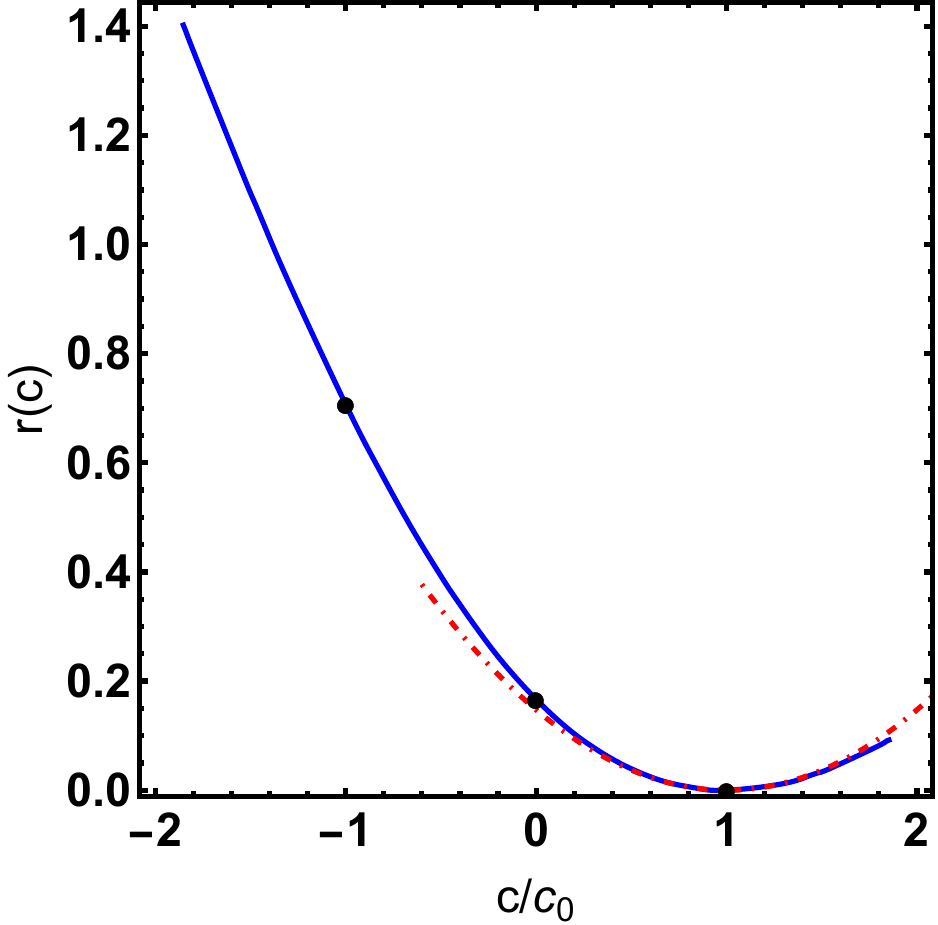}
\caption{The rate function $r(c)$, which describes large deviations of the empirical front speed $c$, see Eq.~(\ref{ldf}), vs. $c/c_0$, where $c_0=1/\sqrt{2}$. Shown are numerical results for  $-c_{\text{cr}}\leq c\leq c_{\text{cr}}$ (the blue line) and the Gaussian asymptotic (\ref{reducedaction}) (the red dash-dotted line). The three fat points show the exact results $r(c=c_0)=0$, $r(c=0)=1/6$, and $r(c=-c_0)=c_0$.}
\label{ffig}
\end{figure}

In this way we were able to obtain sufficiently accurate numerical solution for $c$ less than or equal to a critical value $c_{\text{cr}}\simeq 1.3172$.  For larger $c$ our shooting procedure does not converge, signalling a possible change in the character of the optimal solution. We leave this regime of very large positive deviations, $c>c_{\text{cr}}$ [and, by virtue of the fluctuation relation (\ref{accumrateminusfinal}), of extremely large negative deviations, $c<-c_{\text{cr}}$] of the front speed for a future work \cite{regimechange}.

A plot of the rate function $r(c)$, evaluated on the numerically found TFSc for $-c_{\text{cr}}\leq c\leq c_{\text{cr}}$, is shown in Fig. \ref{ffig}. To increase the number of numerical points, we used the fluctuation theorem (\ref{accumrateminusfinal}).  The three fat points show exact analytical results. The typical fluctuations around $c=c_0$ are Gaussian in perfect agreement  with the predictions from the Langevin equation, presented in Sec. \ref{Df}. As Fig. \ref{ffig} also shows, outside of the Gaussian region the speed fluctuations become sub-Gaussian at $c>c_0$, and super-Gaussian at $c<c_0$.

\section{Monte Carlo simulations}
\label{MC}

\subsection{Method}
\label{method}

We performed MC simulations of the stochastic HZ model on a one-dimensional lattice. The simulations employ the standard Gillespie algorithm \cite{Gillespie} with the hopping rate $D_0$ and the rates of reactions $2A\rightleftarrows 3A$, defined by the on-site  master equation (\ref{onsitemaster}). The instantaneous state of the system realization is determined by $n_j$, where $1\leq j \leq L$ is the number of the site.
First we compute, for each site, the instantaneous local activity rate, which is the sum of all of the onsite rates, $A_{n_j}=n_j D_0+\lambda_{n_j}+\mu_{n_j}$.
Then we randomly choose a site $j$ with a probability proportional to $A_{n_j}$, and a particle on that site which can proliferate with the probability $\lambda_{n_j}/A_{n_j}$,
die with the probability $\mu_{n_j}/A_{n_j}$,
or hop to a neighboring site with the probability $n_jD_0/A_{n_j}$.
Hopping to the left is prohibited for particles on site $j=1$, and we make sure that, during the simulation time, particles do not reach the right edge of the system, $j=L$. After each elementary process, chosen by this algorithm, the time is (on average) advanced by
$\delta t_{n_j} =n_j/(MA_{n_j})$, where $M$ is the total number of particles in the system at this moment of time.

We use the dimensionless  parameters $K$ and $N$ for a ``universal" representation of simulations results for different sets of the original lattice model parameters $\sigma$ and $D_0$ (the parameters $\beta$ and $h$ could be set to one from  the start).  The simulations were performed in the parameter region of the $(K,N)$ plane which obeys the strong inequality~(\ref{NandK}).  For each set of the parameters,  we generated $\mathcal{M}$ independent realizations to form a Monte-Carlo ensemble. Typically, $\mathcal{M}$ varied in the range $200\leq\mathcal{M}\leq2000$.

Particles were initially placed at the left part of the lattice, $x<x_0$, while the right part of the system was unoccupied. As expected, this population of particles start to invade the previously unoccupied sites at $x>x_0$. Eventually a density front forms and propagates to the right. At long times, $t\gg 1$ (we actually had to wait until $t\gtrsim 20-40$) the Monte-Carlo ensemble becomes statistically time-independent.

As an example, Fig.~\ref{figsim0} shows  the density profiles of the propagating front at two different times as observed in a single MC simulation with $K=20$ and $N\simeq 173$. As to be expected, these density profiles agree very well with a numerical solution of the deterministic HZ equation ~(\ref{Z}) and, at late times, with
the TFS presented in Eq.~~(\ref{mffront}).

The  position $X(t)$ of the simulated stochastic front at time $t$ is determined by the realization $n_j$ at this time,  and we defined $X(t)$  as follows. Consider the total number of particles, $\mathcal{N}_j$, on the sites which are to right of the site $j$:
\begin{equation}\label{totheright}
\mathcal{N}_j=\sum_{j^\prime>j}n_{j^{\prime}}.
\end{equation}
The leftmost site $J$, for which $\mathcal{N}_J$ is larger than or equal to $N$, defines the position of front via the expression
$X(t)=J h /\ell_D$. Note that, in the limit of $N\to\infty$ one has $\int_{x>X(t)} u(x,t)\, dx=1$.

Roughly speaking,  this definition attributes the front position $X(t)$ to the $N$-th rightmost particle in the front and is therefore
a good characteristic of macroscopic fluctuations of the front. From the viewpoint of the deterministic TFS~(\ref{mffront}), this definition of $X(t)$ corresponds to a value of $U_0$ which is close to $1/2$: $U_0(X-ct)\simeq 0.5069$.

\subsection{Results}
\label{numresults}

We focused on measuring the mean increment, $\langle X(t+\Delta t) - X(t)\rangle$, the variance $\mbox{Var}\, (X(t+\Delta t) - X(t))$, and
the PDF of the increment of the time-dependent front position $X(t)$ at long times, $t\gg 1$, when a statistical steady-state is observed in the simulations. The mean increment
yields the asymptotic empirical front speed $c_*(N)$, see Sec. \ref{intro}. In its turn, the variance of the increment
yields an \emph{apparent} front diffusion coefficient,
\begin{equation}\label{P140}
D_*(\Delta t;N)=\frac{\mathcal{V}(\Delta t;N)}{2\Delta t}\, ,
\end{equation}
where $\mathcal{V}$ is defined in Eq.~(\ref{functionC}),
whereas $\Delta t \to \infty$ limit of $D_*(\Delta t;N)$ determines the asymptotic front diffusion coefficient,
\begin{equation}\label{Dffinal}
D_f(N)=\lim_{\Delta t\to\infty} D_*(\Delta t; N).
\end{equation}

\begin{figure}[ht]
\includegraphics[width=3.0 in,clip=]{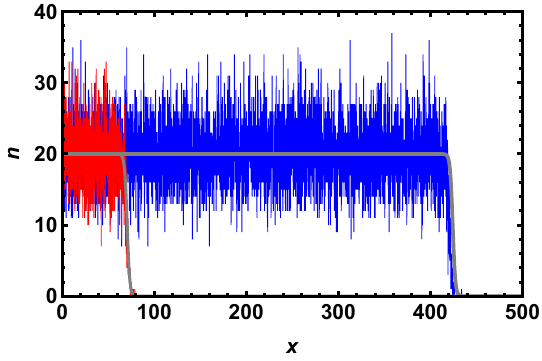}
\caption{The density profiles of a stochastic HZ front at times $t=100$ (red) and $t=600$ (blue),
observed in a single MC simulation. The gray lines show the deterministic TFS~(\ref{mffront}) at the respective times. The rescaled parameters are $K=20$ and $N \simeq 173$.}
\label{figsim0}
\end{figure}

\begin{figure}[ht]
\includegraphics[width=2.5 in,clip=]{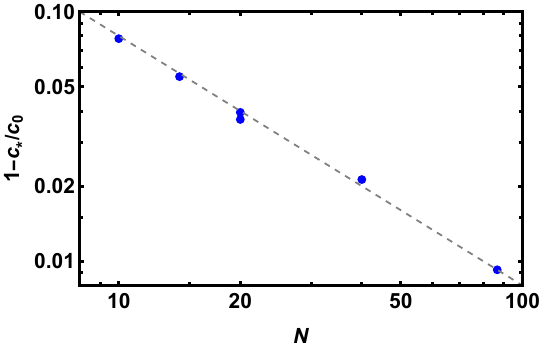}
\caption{Finite-$N$ correction to the average front speed $c_0$. Plotted is the difference $1-c_*(N)/c_0$  vs. $N$. Blue points: the results of six different sets of simulations: $K=1$ and $N=10$, $K=1$ and $N\simeq14$, $K=1$ and $N=20$, $K=2$ and $N=20$,  $K=2$ and $N=40$, and $K=5$ and $N\simeq 87$. Straight line: the fit $c_*(N)=c_0(1-0.8/N)$.
}
\label{fig_delta_c}
\end{figure}

Let us start with the analysis of the measurements of the average front speed $c_*(N)$. As one can see from Fig.~\ref{figsim0}, the average front speed is very close to, but slightly smaller than, the mean-field value $c_0=1/\sqrt{2}$. In Fig.~\ref{fig_delta_c} we plot the observed relative difference, $1-c_*(N)/c_0$ as a function of $N$. Note that the size of the simulated blue points is about twice the standard deviation, obtained from $\mathcal{V}(\Delta t\gg1;N)$. The two closely lying points at $N=20$ correspond to two series of simulations with $K=1$ and $K=2$. Taking into account the statistical uncertainty, we can conclude that the velocity shift mostly depends on $N$, but not on $K$. The dashed line in  Fig.~\ref{fig_delta_c} shows the fit
\begin{equation}\label{P210}
c_*(N)=c_0\left(1-\frac{\alpha_c}{N}\right)\, ,\mbox{~~~where~~~} \alpha_c =0.8\,,
\end{equation}
of the $N$-dependence of $c_*(N)$. As one can see, the $N$-dependent correction to the deterministic front speed of the HZ front is negative. Further, it scales as $1/N$, as to be expected from a second-order perturbation theory in the small parameter $1\sqrt{N}$, applied to the Langevin equation (\ref{P040}).

\begin{figure}[ht]
\includegraphics[width=3.0 in,clip=]{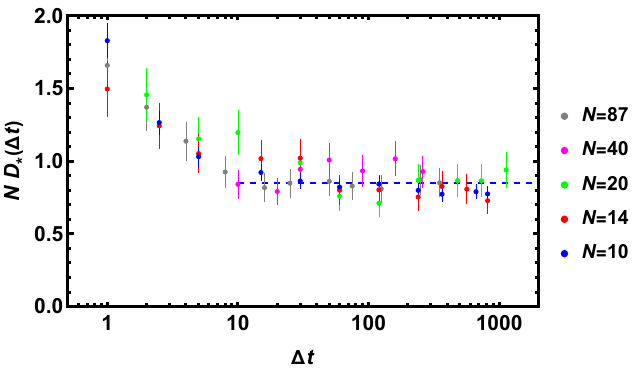}
\caption{Apparent front diffusion coefficient~(\ref{P140}), rescaled by $1/N$, vs. time lag $\Delta t$, for
four sets of simulations: $K=1$,  $N=10$, $\mathcal{M}=2000$ (blue); $K=1$, $N=\sqrt{200}\simeq14$, $\mathcal{M}=200$  (red); $K=2$, $N=20$, $\mathcal{M}=200$  (green); $K=2$, $N=40$, $\mathcal{M}=800$  (magenta); and $K=5$, $N=\sqrt{7500}\simeq 87$, $\mathcal{M}=600$ (gray). The logarithmic scale for $\Delta t$ is used.
The error bars show statistical standard deviations determined by the number of realizations, $\mathcal{M}$, for each combination of the parameters.
}
\label{figNDstar}
\end{figure}

Our simulations results for the (rescaled by $1/N$) apparent diffusion coefficient of the front $D_*(\Delta t,N)$ as a function of the time lag $\Delta t$ are presented in Fig.~\ref{figNDstar}.
The standard-deviation error bars, shown in this plot,
are determined by the number $\mathcal{M}$ of the simulated realizations: $\mbox{Var}\, D_*\simeq 2 D_*^2/\mathcal{M}$.
As one can see, at large $\Delta t$
the function $N D_*(\Delta t; N)$ approaches a constant value $A_*\simeq 0.8-0.9$.
This yields the asymptotic diffusion coefficient of the front,
\begin{equation}\label{P230}
D_f\simeq \frac{A_*}{N}\,.
\end{equation}
which is in agreement with our theoretical predictions~(\ref{Dfpredicted}) and~(\ref{reducedaction}).

Now let us have a closer look at the transient dynamics of the apparent coefficient of front diffusion, as shown in Fig.~\ref{figNDstar}. One can see that, with the accuracy determined by the statistical spread of our data (which we estimate as $3 - 10 \%$), the simulated points collapse into a single curve which is independent of $N$.  Therefore, not only the asymptotic front diffusion coefficient $D_f$, but also the apparent diffusion coefficient $D_*(\Delta t;N)$ scale as $1/N$ (and is independent of $K$). This observation suggests that even the transient behavior of the front at intermediate time lags is describable by the Langevin equation~(\ref{P040}), with its single parameter $N\gg 1$, or in the framework of the full time-dependent MFT, see Sec.~\ref{MFTeq}.

\begin{figure}[ht]
\includegraphics[width=3.0 in,clip=]{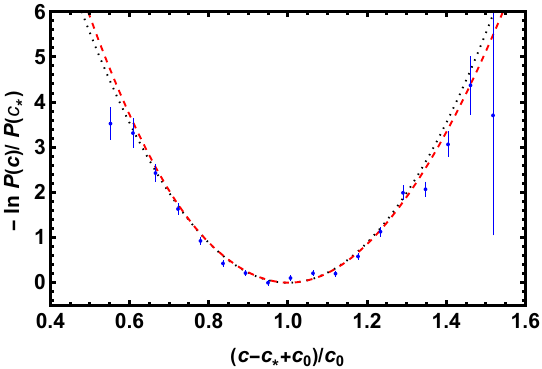}
\caption{The (properly normalized) PDF of the empirical front speed. The blue points with error bars show the simulated $-\ln \left[\mathcal{P} (c)/\mathcal{P} (c_*)\right]$ vs. $(c-c_*-c_0)/c_0$ for $N=10$, $K=1$, $\mathcal{M}=2000$, and $\Delta t=15$.
The dotted black line shows the Gaussian asymptotic~(\ref{reducedaction}), the dashed red line shows the full numerical rate function $r(c)$, presented in Fig.~\ref{ffig}.
}
\label{figPDFsim}
\end{figure}

Our simulation results also allowed us to measure the PDF of the front speed $c$: in a limited range $|c-c_*|\lesssim c_0$ and for not too large $\Delta t$. In the standard MC simulations, used in this work, there is a trade-off between  $\Delta t$ and the range of $c$ available for the analysis. The larger $\Delta t$ is, the closer the PDF is to its asymptotic form that can be compared with the results of Secs.~\ref{Df} and~\ref{WKB}.
On the other hand, too large values of $\Delta t$ make the larger deviations, where $|c-c_*|$ can be of the order of $c_0$, highly improbable and therefore inaccessible in standard simulations.   A compromise between these two contradictory demands is the choice of $\Delta t \simeq 10-20$, where the systematic uncertainty, caused by the finite $\Delta t$, is comparable with the statistical spread caused by the finite sampling size $\mathcal{M}$, see Fig.~\ref{figNDstar}.

The resulting PDF of $c$, presented in Fig.~\ref{figPDFsim}, was obtained for $K=1$, $N=10$ and $\Delta t=15$. This figure also shows our theoretical predictions: the Gaussian asymptotic~(\ref{reducedaction}) and the numerical results from Fig.~\ref{ffig}. As one can see,  at $|c-c_*|\lesssim c_0$ the simulation results agree with the theory quite well. Note that the difference between the exact rate function and its Gaussian asymptotic is quite small in this range of $c$, and the simulations cannot discern it.

Finally, we investigated the sensitivity of the front diffusion measurements to a change in the definition of the front position $X(t)$.
To remind the reader, we measured the front position by, roughly speaking, the position  of the $N$-th rightmost particle.
How will the results change if we measure $X(t)$ by the number of the site containing the \emph{first} particle of the front?
We found out (see the Appendix for details) that,  at $\Delta t=\mathcal{O}(1)$, the apparent front diffusion coefficient $D_*$   is very large for this protocol: $\mathcal{O}(1)$ independently of $N\gg 1$. Importantly, $D_*$ does ultimately approach the asymptotic value $D_f$, but this happens only at extremely long times, $\Delta t\gtrsim t_f(N)$, where $t_f(N)$ grows with $N$ approximately linearly. 
Combining these results with the ones presented above (for our standard protocol), we can conclude that (1) the leading edge of the front fluctuates strongly, (2) the strong fluctuations persist at $N\to\infty$, and (3) the strong fluctuations do not affect the dynamics of the main body of the front.

\section{Discussion}
\label{discussion}

The deterministic HZ equation~(\ref{Z}) describes a special case of a reaction-diffusion equation, where a linear reaction term is absent. Let us temporarily reintroduce this term,
\begin{equation}\label{epsilonterm}
\partial_t u =\epsilon u +u^2 -u^3 +\partial_x^2 u,
\end{equation}
and assume that $|\epsilon| \ll 1$. In the context of reactions, 
the linear term corresponds to the decay $A\to 0$ (for $\epsilon<0$) or to the simple branching $A\to 2A$ (for $\epsilon>0$).

For $\epsilon<0$ the empty state $u=0$ is linearly stable but metastable, and Eq.~(\ref{epsilonterm}) describes invasion of this state by the stable populated state at $u\simeq 1+\epsilon$. In this case the front speed
is determined by the whole set of reactions in the r.h.s. of Eq.~(\ref{epsilonterm}) \cite{vanSaarloos03}. In their turn, typical fluctuations of the corresponding \emph{stochastic} front
exhibit a relatively simple behavior with $N\gg 1$: both the asymptotic correction to the deterministic front speed $\delta c$, and the asymptotic front diffusion coefficient $D_f$ in the moving frame scale as $1/N$ \cite{Panja,MSK,Benguria}. Here the magnitude of typical fluctuations in the bulk of the front scales as $1/\sqrt{N}$ already at $\Delta t \gtrsim 1$.

For $\epsilon>0$ the empty state $u=0$ is linearly unstable, but for  $\epsilon<2$ the invasion front remains pushed \cite{vanSaarloos03}. For sufficiently small positive $\epsilon$, typical fluctuations of the corresponding \emph{stochastic} front show the anomalies which we briefly discussed in Sec. \ref{intro}: (i)  At  $N\to\infty$ and times $\Delta t=\mathcal{O}(1)$, the fluctuations in the bulk of the front are $\mathcal{O}(1)$, and (ii) the front exhibits a diffusion-type behavior, with $D_f\sim D/N$ scaling, only
when measured at extremely large time lags $\Delta t\gg N\gg 1$ \cite{KMS}.

The HZ front that we studied here is exactly at the borderline $\epsilon=0$ between the regimes of $\epsilon<0$ and $\epsilon>0$.   Therefore, it is not obvious a priori how the stochastic HZ front behaves: in particular, whether or not it shows the intermediate-time anomaly.  Essentially, our MC simulations, presented in Sec. \ref{MC}, have shown that the stochastic HZ front case behaves in the same way as the metastable fronts with their (i) ``standard" $1/N$ scaling behavior of $\delta c$ and $D_f$, (ii) a relatively fast convergence to this scaling regime, which only requires the strong inequality $\Delta t \gg 1$. This signals that the leading particles of the front do not play an important role in spite of their very fast diffusion.  In addition, the MFT calculations of Sec. \ref{WKB} show that the large deviations of the stochastic HZ front exhibit a qualitatively similar behavior at $c<c_0$ and $c>c_0$, also negating any important role of the leading particles in the behavior of the bulk of the front.  Still, it would be interesting to pinpoint the mechanism behind large fluctuations of leading particles at moderate time lags.

Finally, it would be interesting to investigate the regime of extremely large deviations of the front speed, $|c|>c_{\text{cr}}\simeq 1.3172$, where the optimal history of the conditioned process, and the ensuing probability density $\mathcal{P}(c,\Delta t,N)$, are yet unknown.

\textbf{Acknowledgments.}  B. M. was supported by the Israel
Science Foundation (Grant No. 1579/25).

\appendix

\section*{Appendix. Fluctuations of the rightmost particle}
\label{AppA}

Here we present some details concerning the variance $\mathcal{V}_1$ of the increment of the position of the first particle of the front (more precisely, of the number of the rightmost site containing at least one particle). This quantity is defined similarly to $\mathcal{V}$  in Eq.~(\ref{functionC}), except that the position of the front $X(t)$ is replaced by the  position of the first particle, $X_1(t)$. We focused on the apparent diffusion coefficient of the first particle,
\begin{equation}\label{A010}
D_{*\,1}(\Delta t;N)=\frac{\mathcal{V}_1(\Delta t;N)}{2\Delta t}\,,
\end{equation}
defined by analogy with D$_{*}(\Delta t;N)$ from Eq.~(\ref{P140}).

\begin{figure}[ht]
\includegraphics[width=3.0 in,clip=]{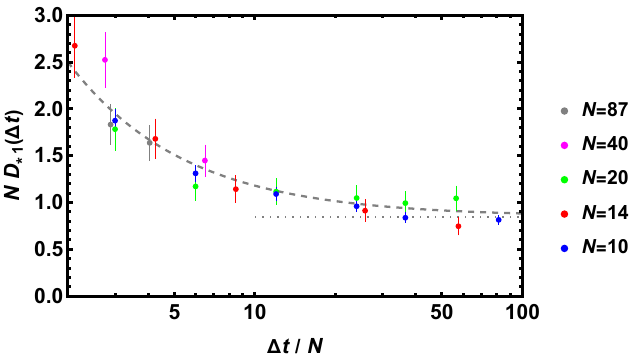}
\caption{Rescaled diffusion coefficient of the first particle $ND_{*\,1}(\Delta t; N)$ versus rescaled time lag $\Delta t/N$ for the simulations with the following parameters: $K=1$, $N=10$, $\mathcal{M}=2000$ (blue); $K=1$, $N=\sqrt{200}\simeq14$, $\mathcal{M}=200$ (red); $K=2$, $N=20$, $\mathcal{M}=200$ (green); $K=2$, $N=40$, $\mathcal{M}=800$ (magenta); and $K=5$, $N=\sqrt{7500}\simeq 87$, $\mathcal{M}=600$ (gray). The dotted line shows the theoretical result~(\ref{Dfpredicted}), and the dashed line shows the empirical fit~(\ref{A020}).}
\label{figNDstar1}
\end{figure}

Our first observation is that the behavior of $D_{*\,1}(\Delta t;N)$ at  $\Delta t\to \infty$ agrees fairly well with the theoretical prediction~(\ref{Dfpredicted}) following from theory~\cite{MSK}.  Figure~\ref{figNDstar1} shows $ND_{*\, 1} (\Delta t)$, extracted from the simulations, versus $\Delta t/N$ for different values of $N$. The dotted line shows the theoretical limit  $ND_f$. This figure also shows an empirical fit of the data:
\begin{equation}\label{A020}
D_{*\,1}(\Delta t;N)\simeq\frac{0.85}{N}+\frac{3.3}{\Delta t}
\quad
(N\gg1\mbox{~and~} \Delta t\gg N)\, .
\end{equation}
This empirical fit is convenient, because it allows one to see that, for the rightmost particle, the theoretical asymptotic value $D_f$ is reached  only at extremely long time lags, $\Delta t \gtrsim 30 N$.

\begin{figure}[ht]
\includegraphics[width=3.0 in,clip=]{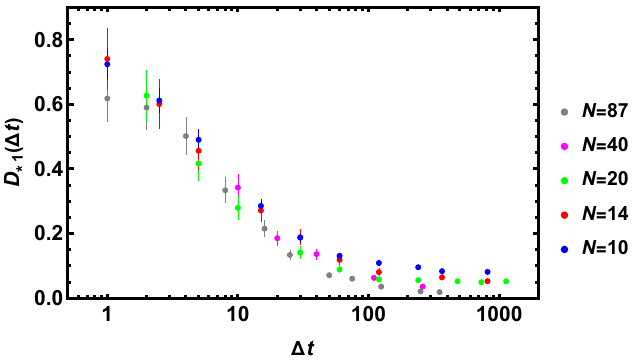}
\caption{$D_{*\,1}(\Delta t; N)$ versus $\Delta t$ for the simulations with parameters $K=1$, $N=10$, $\mathcal{M}=2000$ (blue); $K=1$, $N=\sqrt{200}\simeq14$, $\mathcal{M}=200$ (red); $K=2$, $N=20$, $\mathcal{M}=200$ (green); $K=2$, $N=40$, $\mathcal{M}=800$ (magenta); and $K=5$, $N=\sqrt{7500}\simeq 87$, $\mathcal{M}=600$ (gray).}
\label{figDstar1}
\end{figure}

More surprising results for the first-particle measurement protocol are observed at relatively small time lags, $\Delta t \sim 1-10$. Figure~\ref{figDstar1} presents  $D_{*\, 1}(\Delta t,N)$ versus $\Delta t$ (neither of which are rescaled with $N$) for different $N$. As one can see, at $\Delta t\lesssim 10$, the simulated points with different $N$ collapse into a single curve  which does not depend on $N$. The apparent $N$-independence of (large) fluctuations of the speed of the first particle stands in a striking contrast with the results for the $N$-th particle protocol,  presented in the main text, see Fig.~\ref{figNDstar}. 


Overall, the fluctuations of the leading particle are dominant at small time lags $\Delta t$ (explaining the collapse of the $D_{*\, 1}(\Delta t,N)$ curves in Fig.~\ref{figDstar1}) and slowly decrease with $\Delta t$. For very large time lags, they become smaller than the ``macroscopic" theoretical $1/\sqrt{N}$ fluctuations [see the empirical fit in Eq.~(\ref{A020})] and the macroscopic predictions are recovered as shown in Fig.~\ref{figNDstar1}.

We also measured the empirical mean front speed $\bar{c}_1(N)$ at $t\gg1$, defined by the position $X_1(t)$ of the first particle. 
We observed that  $\bar{c}_1(N)$ does not differ significantly from  the mean front speed $\bar{c}(N)$ considered in the main text.


\begin{thebibliography}{99}
\bibitem{macroscopic} The macroscopic description is expected to hold on spatial scales much larger than the lattice spacing, and on time scales much longer than the typical hopping time $1/D_0$ between the neighboring lattice sites.
\bibitem{discreteness} An additional, purely deterministic, mechanism of a systematic shift of the front speed is due to
the lattice discreteness \cite{Keener}. This shift can be calculated by accounting for the next-order term, proportional to $\partial_x^{(4)} u$, in the continuous approximation of the random walk's discrete Laplacian. Then the corresponding small shift of the front speed can be calculated perturbatively \cite{Keener}. For our HZ model, this standard calculation yields
\begin{equation}\label{Keenershift}
\Delta c   = - (1/240\sqrt{2}) (K/N)^2,
\end{equation}
with a very small numerical coefficient. This shift is negilible in comparison with the fluctuation-induced shift $\delta_c = O(1/N)$ as soon as
$N\gg a K^2$, with a very small $a$.  This strong inequality was fulfilled in our MC simulations with a large margin.
\end{thebibliography}
\end{document}